\documentclass[aps,prl,superscriptaddress,twocolumn]{revtex4} 
\usepackage{graphicx} 
\begin{document}

\title{NEW BOUNDS ON THE CKM MATRIX\\
FROM $B \to K\pi\pi$ DALITZ PLOT ANALYSES}

\author{M.~Ciuchini}
\affiliation{Dipartimento di Fisica, Universit\`a di Roma Tre 
and INFN, Sezione di Roma III, Via della Vasca Navale 84, I-00146
Roma, Italy}
\author{M.~Pierini}
\affiliation{Department of Physics, University of Wisconsin, Madison,
  WI 53706, USA}
\author{L.~Silvestrini}
\affiliation{Dipartimento di Fisica, Universit\`a di Roma ``La
  Sapienza''  and INFN, 
  Sezione di Roma, P.le A. Moro 2, I-00185 Rome, Italy}

\begin{abstract}
  We present a new technique to extract information on the Unitarity
  Triangle from the study of $B \to K\pi\pi$
  Dalitz plots.  Using the sensitivity of Dalitz analyses to the
  absolute values and the phases of decay amplitudes and isospin
  symmetry, we obtain a new constraint on the elements of the CKM matrix.
  We discuss in detail the role of electroweak penguins and outline
  future prospects.
\end{abstract}

\maketitle

\section{Introduction}
\label{sec:intro}

The study of flavour and CP violation in $B$ decays allows us to test
the flavour structure of the Standard Model (SM) and to look for New
Physics (NP). In the last few years, $B$ factories have provided us
with a large amount of new data in this field, in particular the first
measurements of the angles $\alpha$ and $\gamma$ of the Unitarity
Triangle (UT). These measurements are in agreement with the indirect
determinations of the UT angles~\cite{utfit,ckmfit} and therefore
provide an important test of the Standard Model and stringent
constraints on NP~\cite{utnp}. 

$B \to K \pi$ decays could in principle constitute another source of
information on $\gamma$ \cite{minchiategammakpi,minchiatesu3ew}.
However, the fact that the tree-level $b \to u$ transition, carrying
the phase $\gamma$, is doubly Cabibbo suppressed in these channels,
together with a large dynamical enhancement of the Cabibbo allowed
penguin contribution \cite{lavorifondcharming}, make a
model-independent extraction of $\gamma$ from $B \to K \pi$
impossible. Model-dependent studies of these channels have tackled
this issue with no success \cite{QCDfact,charmingnew}.

In this letter, we consider the possibility of obtaining a tree-level
determination of $\gamma$ from $B \to K^* \pi$ decays, cancelling the
effect of penguins through the rich set of information available from
Dalitz plot analyses. In particular, the $B$ factories can provide the
magnitude and phase of decay amplitudes separately for $B$ and $\bar
B$ decays. One can think of exploiting this information, together with
isospin symmetry, to build combinations of amplitudes that are
proportional to a single weak phase.

This very simple idea, however, in the case of $B\to K\pi\pi$, has to face
the presence of Electroweak Penguins (EWP), doubly Cabibbo enhanced
with respect to current-current operators in $b \to s$ transitions.
This enhancement can largely compensate the $\alpha_{em}$ suppression,
leading to an $\mathcal{O}(1)$ correction to the decay amplitude to
the $I = 3/2$ final state. This fact, however, does not spoil the
possibility of extracting information on the UT with small hadronic
uncertainty, as explained below.

Dalitz plot analyses combined with isospin have already shown
their effectiveness in the extraction of $\alpha$ from
$B\to\pi\pi\pi$~\cite{rhopi}. A proposal relying on isospin for extracting
$\gamma$ using a global analysis of $B\to K_S (\pi\pi)_{I=0,2}$, including
time-dependent CP asymmetries at fixed values of the Mandelstam variables,
can be found in ref.~\cite{deshpande}. We focus instead on $K^*\pi$ final
states which allow us to perform a simpler analysis with no need of
time-dependent measurements.

\section{Extracting CKM matrix elements from $B \to K \pi\pi$ Dalitz
  plots}
\label{sec:kpp}

Let us first illustrate our idea for the simplified case in which we
neglect EWP contributions. To this aim, we write the amplitudes 
using isospin symmetry, in terms of 
Renormalization Group Invariant (RGI) complex parameters~\cite{burassilv},
obtaining
\begin{eqnarray}
  A(K^{*+} \pi^-)           &=& ~~V_{tb}^* V_{ts} P_1 - V_{ub}^*
  V_{us} (E_1 - P_1^\mathrm{GIM})  \nonumber\\
  {\sqrt{2}}A(K^{*0} \pi^0) &=& -V_{tb}^* V_{ts} P_1 - V_{ub}^*
  V_{us} (E_2 + P_1^\mathrm{GIM}) \nonumber \\
  {\sqrt{2}}A(K^{*+} \pi^0) &=& ~~V_{tb}^* V_{ts} P_1
  \\ &&- V_{ub}^*
  V_{us} (E_1 + E_2 + A_1 -P_1^\mathrm{GIM}) \nonumber \\
  A(K^{*0} \pi^+)           &=& -V_{tb}^* V_{ts} P_1 + V_{ub}^*
  V_{us} (A_1 - P_1^\mathrm{GIM}), \nonumber
\label{eq:amp}
\end{eqnarray}
where $P_1^\mathrm{(GIM)}$ represent (GIM-suppressed) penguin
contributions, $A_1$ the disconnected annihilation and $E_1$ ($E_2$)
the connected (disconnected) emission topologies. $\bar B$ decay
amplitudes are simply obtained by conjugating the CKM factors
$V_{ij}$. Similar expressions hold for higher $K^*$ resonances.

Considering the two combinations of amplitudes
\begin{eqnarray}
A^0&=&A(K^{*+}\pi^-)+{\sqrt{2}}A(K^{*0} \pi^0)\nonumber\\
&=&-V_{ub}^* V_{us} (E_1 +E_2)\,,\label{eq:atn}\\
\bar A^0&=&A(K^{*-}\pi^+)+{\sqrt{2}}A(\bar K^{*0} \pi^0)\nonumber\\
&=&-V_{ub} V_{us}^* (E_1 +E_2)\,,\label{eq:atnb}
\end{eqnarray}
the ratio
\begin{equation}\label{eq:rn}
R^0=\frac{\bar A^0}{A^0}=\frac{V_{ub} V_{us}^*}{V_{ub}^* V_{us}}=e^{-2 i\gamma}
\end{equation}
provides a clean determination of the weak phase $\gamma$.  We now
discuss how to extract $A_0$ and $\bar A_0$. Looking at the decay
chains $B^0 \to K^{*+}(\to K^+\pi^0)\pi^-$ and $B^0 \to K^{*0}(\to
K^+\pi^-)\pi^0$, one can obtain $A^0$ from the $K^+\pi^-\pi^0$ Dalitz
plot, including the phase in a given convention, for example
Im$A(K^{*+}\pi^-)=0$. Similarly, $\bar A^0$ can be extracted from the
$K^-\pi^+\pi^0$ Dalitz plot using the same procedure, choosing
Im$A(K^{*-}\pi^+)=0$.  However, in general, this choice does not
reproduce the physical phase difference between $A(K^{*+}\pi^-)$ and
$A(K^{*-}\pi^+)$, which has to be fixed using additional information.

This information can be provided by the $K_S\pi^+\pi^-$ Dalitz plot,
considering the decay chain $B^0 \to K^{*+}(\to K^0\pi^+)\pi^-$ and
the CP conjugate $\bar B^0 \to K^{*-}(\to \bar K^0\pi^-)\pi^+$. These
two decay channels do not interfere directly on the Dalitz plot, but
they both interfere with the decays $B,\bar B\to \rho^0 (\to
\pi^+\pi^-) K_S$ and with other resonances contributing to the same
Dalitz plot. Therefore the Dalitz analysis of $B,\bar B\to
K_S\pi^+\pi^-$ should include the $\rho^0 K_S$ final state.  In a
time-integrated analysis, the $\rho^0 K_S$ final state comes from a
mixture of $B$ and $\bar B$ mesons, while the $K^{*+(-)}\pi^{-(+)}$
final state only originates from $B$ ($\bar B$) decay. Looking at the
phases of $A(K^{*+(-)}\pi^{-(+)})$ relative to $A(\rho^0 K_S)$, we can
extract the phase difference between $A(K^{*+}\pi^-)$ and
$A(K^{*-}\pi^+)$ with no need of resolving the flavor of the $B$ in
$A(\rho^0 K_S)$~\footnote{We neglect direct CP violation in $B,\bar
  B\to \rho^0 (\to \pi^+\pi^-) K_S$ decays. Alternatively, one could
  use interference with $B,\bar B\to \chi_c (\to \pi^+\pi^-) K_S$
  decays or the time-dependent $K_S \pi^0 \pi^0$ Dalitz plot to fix
  the phase difference between the $K^{*0}\pi^0$ and $\bar K^{*0}
  \pi^0$ amplitudes. In this case, however, one has to take into
  account the $B$--$\bar B$ mixing phase $2 \beta$.}.

A similar isospin relation involves charged $B$ decays. We have
\begin{eqnarray}
A^+&=&A(K^{*0}\pi^+)+{\sqrt{2}}A(K^{*+} \pi^0)\nonumber\\
&=&-V_{ub}^* V_{us} (E_1 +E_2)\,,\label{eq:atc}\\
A^-&=&A(\bar K^{*0}\pi^-)+{\sqrt{2}}A(K^{*-} \pi^0)\nonumber\\
&=&-V_{ub} V_{us}^* (E_1 +E_2)\,,\label{eq:atcb}
\end{eqnarray}
and the ratio
\begin{equation}\label{eq:rc}
R^\mp=\frac{A^-}{A^+}=e^{-2 i\gamma}\,.
\end{equation}
As before, $A^\pm$ can be extracted from the decay chains $B^\pm \to
K^{*\pm}(\to K^0\pi^\pm)\pi^0$ and $B^\pm \to K^{*0}(\to
K^0\pi^0)\pi^\pm$ entering the $K_S\pi^\pm\pi^0$ Dalitz plot.
Electric charge forbids the extraction of the relative phase of the
two Dalitz plots along the way discussed above, so that a strategy
based on theoretical arguments has to be adopted.  In particular, one
can follow two possible paths.

The first one is to use isospin symmetry to relate charged and neutral
$B$ decays:
\begin{eqnarray}
  \label{eq:relations1}
  &&A(K^{*+} \pi^-)+{\sqrt{2}}A(K^{*0} \pi^0)\nonumber\\
  &&- A(K^{*0}
  \pi^+)-{\sqrt{2}}A(K^{*+} \pi^0) =0 \\
  &&A(K^{*-} \pi^+)+{\sqrt{2}}A(\bar K^{*0}
  \pi^0)\nonumber\\
  &&-A(\bar K^{*0} \pi^-)-{\sqrt{2}}A(K^{*-} \pi^0)=0
\end{eqnarray}
so that the relative phases can be fixed. In this way, no additional
information on $\gamma$ can be extracted from $R^\mp$. However, the
full information coming from charged and neutral $B$'s can be combined
to improve the accuracy of the determination of $\gamma$, thanks to
the increase of available statistics.  The second possibility is to
use the penguin-dominated channel $K^{*0} \pi^+$ to fix the phase
difference between the amplitudes in the two Dalitz plots. In this way
an independent, albeit more uncertain, determination of $\gamma$ can
be obtained from $R^\mp$. To illustrate this point, let us write down
the phase for the $K^{*0} \pi^\pm$ final state
\begin{eqnarray}
  \arg \left(A(K^{*0} \pi^+)\right)&=&  \beta_s+
  \arg\left( 1+\frac{V_{ub}^* 
      V_{us}}{V_{tb}^* V_{ts}} \Delta^+ e^{i
        \delta_{\Delta^+}}\right)\nonumber\\
  \arg \left(A(\bar K^{*0} \pi^-)\right)&=&-\beta_s +
  \arg\left( 1+\frac{V_{ub} 
      V_{us}^*}{V_{tb} V_{ts}^*} \Delta^+ e^{i
        \delta_{\Delta^+}}\right)\nonumber
\end{eqnarray}
where $\beta_s=\arg(-V_{ts}V_{tb}^*/(V_{cs}V_{cb}^*))$ and
\begin{eqnarray}
  \label{eq:deltas}
  \Delta^+ e^{i
        \delta_{\Delta^+}} = \frac{A_1 - P_1^\mathrm{GIM}}{P_1}\,.
\end{eqnarray}
We now take advantage of the fact that $|V_{ub}^* V_{us}|/|V_{tb}^*
V_{ts}|\ll 1$ to simplify the above equations and we obtain
\begin{eqnarray}
\arg \left(A(\bar K^{*0} \pi^-)\right) &=& 
  \arg \left(A(K^{*0} \pi^+)\right)
   - 2 \beta_s \nonumber \\ && + 2
  \Delta^+\mathrm{Im} \frac{V_{ub}  
    V_{us}^*}{V_{tb} V_{ts}^*}\cos \delta_{\Delta^+}\,.
  \label{eq:deltaphi}
\end{eqnarray}
On general grounds, we expect $\Delta^+ \sim \mathcal{O}(1)$. The
error induced by the last term in eq.~(\ref{eq:deltaphi}) can be
estimated at the level of $|V_{ub} V_{us}^*|/|V_{tb} V_{ts}^*| \sim
\lambda^2$.
The determination of $\gamma$ from $R^\mp$ is not as theoretically
clean as the one obtained from $R^0$.
Nevertheless, the uncertainty induced by our
dynamical assumption, being of $\mathcal{O}(\lambda^2)$, is much
smaller than the expected experimental error (at least in the near
future).

\section{Inclusion of Electroweak Penguins}
\label{sec:ew}

The inclusion of the effect of EWP's completely changes 
eqs.~(\ref{eq:rn}) and (\ref{eq:rc}).
In fact, even though EPW's give a subdominant contribution to
branching ratios (because of the $\mathcal{O}(\alpha_{em})$ suppression
with respect to the strong penguin contribution), they provide an
$\mathcal{O}(1)$ contribution to $R^0$ and $R^\mp$
(more generally, they provide an $\mathcal{O}(1)$ correction to CP
violating effects in charmless $b \to s$ decays). Fortunately, as we
shall discuss in the following, the dominant EWP's
(\textit{i.e.}~left-handed EWP operators) can be eliminated at
 the operator level, so that no additional hadronic matrix elements are
introduced. The net effect of EWP's is that $R^0$ and
$R^\mp$ depend not only on $\gamma$ but also on other CKM
parameters. 

\begin{figure*}[t]
\caption{Bounds in the $\bar\rho$--$\bar\eta$ plane from
the analysis of $B^0$ decays, assuming a measurement of the
relative phase with an error of $20^\circ$ (left) or $40^\circ$ (right).
The output of the present UT fit analysis~\cite{utfit} is shown as a
reference.}
\label{fig:rhoeta}
\begin{tabular}{l r}
\includegraphics[width=0.45\textwidth]{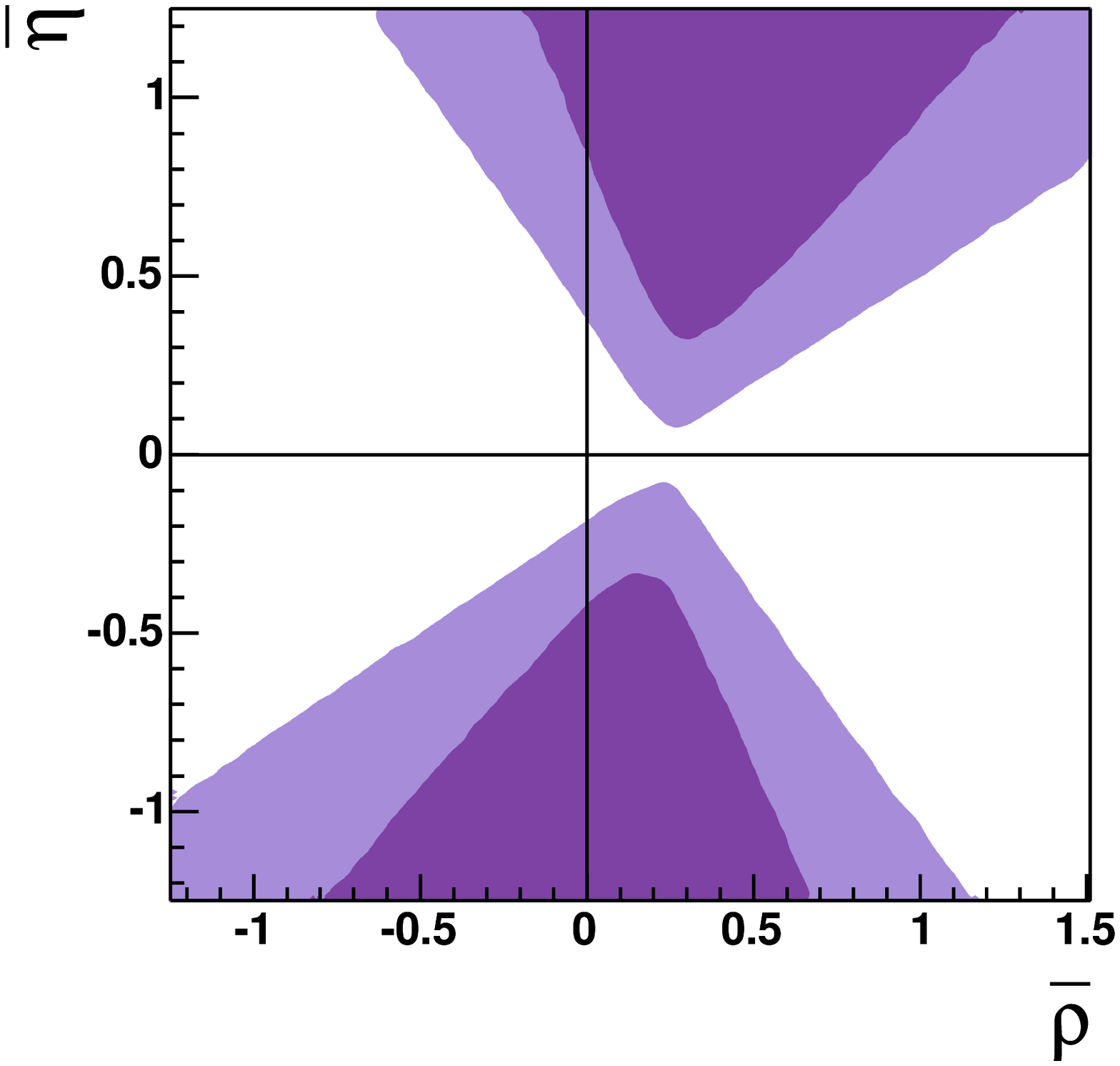} & 
\includegraphics[width=0.45\textwidth]{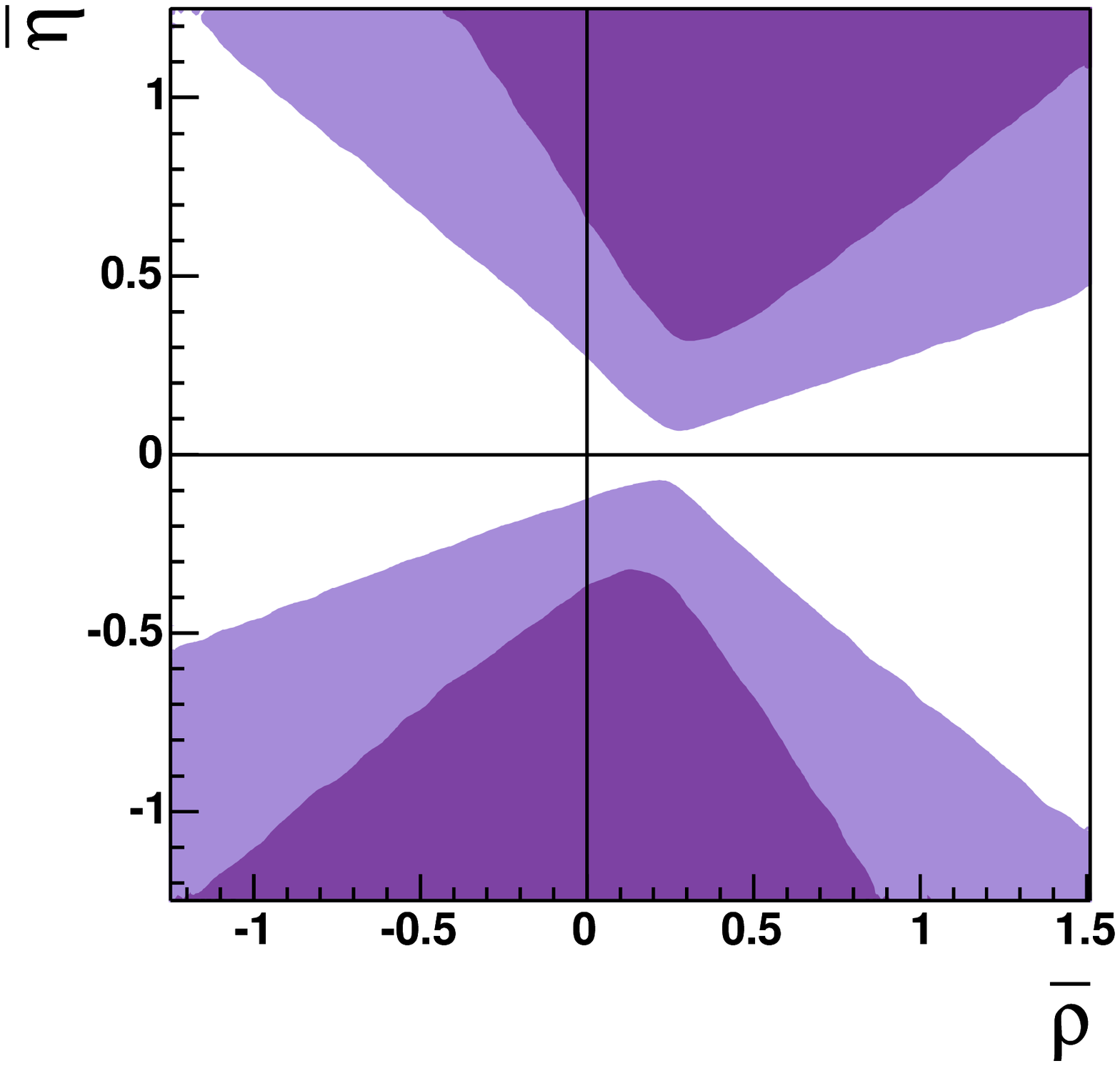}
\end{tabular}
\end{figure*}

Let us consider the effective Hamiltonian for $b \to s$ transitions
given for instance in eq.~(5) of ref.~\cite{burassilv}. There is a
hierarchy in the values of the Wilson coefficients for EWP operators:
$|C_{9,10}| \gg |C_{7,8}|$. Let us therefore neglect the
effect of $Q_{7,8}$ and focus on $Q_{9,10}$, barring a very large
dynamical enhancement of $\langle Q_{7,8} \rangle$ in the case of $B$
decays. There is an exact operator relation that allows to eliminate
$Q_{9,10}$~\cite{anatomy} in favour of current-current and penguin
operators~\footnote{In the literature, it is sometimes claimed
 that $SU(3)$ is necessary to eliminate the matrix elements of
 electroweak penguins. In fact, there is no need to invoke flavour
 symmetries.}:
\begin{eqnarray}
  \label{eq:eliminateQ}
  Q_9 &=& \frac{3}{2} \left(Q_2^{suu} - Q_2^{scc}\right)+3
    Q_2^{scc}-\frac{1}{2} Q_3^s \\
  Q_{10} &=& \frac{3}{2} \left(Q_1^{suu} - Q_1^{scc}\right)+3
    Q_1^{scc}-\frac{1}{2} Q_4^s \nonumber
\end{eqnarray}
so that the effective Hamiltonian becomes
\begin{eqnarray}
  H_\mathrm{eff} &=& \frac{G_F}{\sqrt{2}}\Biggl[
 \left( V_{ub}^* V_{us} C_+-\frac{3}{2} V_{tb}^* V_{ts} C_+^\mathrm{\scriptscriptstyle EW}\right)
    \left(Q_+^{suu} - Q_+^{scc}\right)\nonumber\\ && ~
   +\left( V_{ub}^* V_{us}C_-  +\frac{3}{2} V_{tb}^* V_{ts} C_-^\mathrm{\scriptscriptstyle EW}\right)\left(Q_-^{suu} -
      Q_-^{scc}\right)\nonumber\\&& \quad
   -V_{tb}^* V_{ts} H^{\Delta  I=0}\Biggr]\,.\label{eq:heff}
\end{eqnarray}
where $Q_\pm=(Q_1\pm Q_2)/2$, $C_\pm=C_1\pm C_2$ and $C_\pm^\mathrm{\scriptscriptstyle EW}=C_9\pm C_{10}$ .
We observe that the relation
\begin{equation}
\frac{C_+^\mathrm{\scriptscriptstyle EW}}{C_+}=\frac{C_-^\mathrm{\scriptscriptstyle EW}}{C_-}
\label{eq:ewpc}
\end{equation}
is exact at the LO and is broken by
$\mathcal{O}(\alpha_s\alpha_{em}\log)$ corrections. Numerically,
eq.~(\ref{eq:ewpc}) holds with high accuracy, being violated at the
percent level.  Using this relation, we can write
\begin{eqnarray}
 && H_\mathrm{eff} \simeq \frac{G_F}{\sqrt{2}}\Biggl\{V_{ub}^* V_{us} \left(1+ \kappa_\mathrm{\scriptscriptstyle EW}\right)
    \Bigl[C_+ \left(Q_+^{suu} - Q_+^{scc}\right)+ \nonumber\\ && \quad
    \frac{1-\kappa_\mathrm{\scriptscriptstyle EW}}{1+\kappa_\mathrm{\scriptscriptstyle EW}}
    C_-\left(Q_-^{suu} - Q_-^{scc}\right)\Bigr]-
    V_{tb}^* V_{ts} H^{\Delta  I=0}\Biggr\}\,,\label{eq:heffapp}
\end{eqnarray}
where
\begin{eqnarray}
  \kappa_\mathrm{\scriptscriptstyle EW}&\equiv& - \frac{3}{2} \frac{C_+^\mathrm{\scriptscriptstyle EW}}{C_+}
  \frac{V_{tb}^* V_{ts}}{V_{ub}^* V_{us}}
  \nonumber \\ 
  &=& \frac{3}{2}
  \frac{C_+^\mathrm{\scriptscriptstyle EW}}{C_+}
  \left(1 + \frac{1-\lambda^2}{\lambda^2 \left(\bar
        \rho + i \bar \eta\right)}+ \mathcal{O}(\lambda^2)\right)\,.
  \label{eq:kappa}
\end{eqnarray}
Therefore $R^0$ and $R^\mp$ can be rewritten as
\begin{equation}
  \label{eq:gammaeff}
  R^{0,\mp}=e^{-2 i \left(\gamma+\arg (1+\kappa_\mathrm{\scriptscriptstyle EW})\right)}
\frac{1+\frac{1-\kappa_\mathrm{\scriptscriptstyle EW}^*}{1+\kappa_\mathrm{\scriptscriptstyle EW}^*}\frac{C_-}{C_+} r
  e^{i\theta_r}}
  {1+\frac{1-\kappa_\mathrm{\scriptscriptstyle EW}}{1+\kappa_\mathrm{\scriptscriptstyle EW}}\frac{C_-}{C_+} r 
  e^{i\theta_r}}\,,
\end{equation}
where
\begin{equation}
r e^{i\theta_r}=\frac{\langle K^*\pi (I=3/2)| Q_-|B\rangle}{\langle K^*\pi (I=3/2)| Q_+|B\rangle}\,.
\end{equation}
While for $B\to K\pi$ decays the $SU(3)$ flavour symmetry guarantees
that $\langle K\pi (I=3/2)| Q_-|B\rangle$
vanishes~\cite{minchiatesu3ew,sanda}, the same argument, based on the
symmetry property of the final state wave function, does not apply to
$K^*\pi$ final states~\cite{gronau}.  However, using factorized
amplitudes and form factors as given in ref.~\cite{QCDfact}, one
obtains
\begin{equation}
r=\left\vert\frac{f_{K^*} F_0^{B\to \pi}-f_\pi A_0^{B\to K^*}}
{f_{K^*} F_0^{B\to \pi}+f_\pi A_0^{B\to K^*}}\right\vert \lesssim 0.05\,.
\end{equation}
While this numerical result depends on the estimate of the form
factors, the good agreement between QCD sum rules and lattice QCD
calculations makes it rather robust. In our analysis, however, we do
not assume a specific model for computing the amplitudes, rather we
let $r$ to vary in the conservative range $0$--$0.3$.

Concerning $\kappa_\mathrm{\scriptscriptstyle EW}$, using the values
$C_+(m_b)=0.877$, $C_+^\mathrm{\scriptscriptstyle EW}(m_b)=-1.017\,
\alpha_{em}$ and $\bar\rho = 0.216$, $\bar\eta = 0.342$~\cite{utfit},
we obtain $\kappa_\mathrm{\scriptscriptstyle EW}=-0.35+0.53\, i$.  One
can thus verify that $\kappa_\mathrm{\scriptscriptstyle EW}$ is an
$\mathcal{O}(1)$ correction to the decay amplitude to the $I=3/2$
final state.

The above equations allow to translate the experimental results for
$R^0$ and $R^\mp$ into allowed regions in the $\bar\rho$--$\bar\eta$ plane.
Neglecting terms of $\mathcal{O}(r^2)$, for a given value of $R^0$ one
obtains the linear relation $\bar\eta=-\tan(\frac{1}{2}\arg
R^0)(\bar\rho-\bar\rho_0)$, where
\begin{eqnarray}
  \bar\rho_0 &=& -\left[
    \frac{3C_+^\mathrm{\scriptscriptstyle EW}}{ 2C_+ + 
      3 C_+^\mathrm{\scriptscriptstyle EW}}  - \frac{
      12C_+^\mathrm{\scriptscriptstyle EW} C_- r\cos \theta_r}{\left( 2C_+ + 
        3 C_+^\mathrm{\scriptscriptstyle EW} \right)^2}\right] \cdot \nonumber \\
  && \quad \frac{1 - \lambda^2}{\lambda^2} + \mathcal{O}(\lambda^2)
\label{eq:linearized}
\end{eqnarray}
Including terms of $\mathcal{O}(r^2)$ or higher, the relation between
$\bar \eta$ and $\bar \rho$ becomes quadratic, but the deviation from
the result in eq.~(\ref{eq:linearized}) is small and mainly amounts to
a depletion of the region $\bar \eta \sim 0$. Furthermore, the effect of
$r$ is constrained by the measurement of $\vert R^{0,\mp}\vert$,
since$\vert R^{0,\mp}\vert-1 \propto r \sin \theta_r$.

We can test this new idea using the experimental result of the BaBar
Collaboration on $B^0 \to K^+\pi^-\pi^0$ decays~\cite{babarkpp0}.
Among the other measurements, this analysis provides the decay
amplitudes for $B^0$ and $\bar B^0$ decays to $K^*\pi$ and to
$K^*(1430)\pi$ final states.  We implement this information in our
analysis using directly the shape of the multidimensional likelihood
from BaBar (including all correlations).  In this way, we obtain an
error of $38^\circ$ on $\arg R^0$. Unfortunately, the $K_S\pi^+\pi^-$
Dalitz plot is not yet available so that we cannot fix the relative
phase of $B$ and $\bar B$ decays at present.  For the sake of
illustration, we assume a central value for this relative phase such
that the constraint on $\bar\rho$ and $ \bar\eta$ from
eq.~(\ref{eq:gammaeff}) is compatible with the SM UT fit result
\cite{utfit}.  For the experimental uncertainty on the relative phase,
we consider two cases, corresponding to $\pm 20^\circ$ or $\pm
40^\circ$, leading to the constraints exhibited in
Fig.~\ref{fig:rhoeta}, obtained without expanding in $r$.

The situation can be improved by fitting $R^0_{K^*\pi}$ and
$R^0_{K^*(1430)\pi}$ directly from data, cancelling out part of the
systematic error. In addition, the determination of the UT parameters
can be further improved with the experimental measurement of
$R^\mp_{K^*\pi}$ and $R^\mp_{K^*(1430)\pi}$ and adding Belle data.

Let us finally comment on the sensitivity to NP of our analysis.
Making the very reasonable assumption that NP effects only enter at
the loop level, we can envisage three possibilities. First of all, NP
could affect the coefficients of QCD penguin operators. In this case,
the analysis of $R^0$ is completely unaffected, while the phase of NP
contributions would modify eq.~(\ref{eq:deltaphi}). This could produce
a discrepancy between the constraints on the UT obtained from $R^0$
and $R^\mp$ using eq.~(\ref{eq:deltaphi}). A second possibility is
that NP modifies EWP coefficients, respecting however the hierarchy
$C_{9,10}\gg C_{7,8}$. In this case, the only effect would be a
modification of $\kappa_\mathrm{\scriptscriptstyle EW}$, so that the
constraint on the UT obtained using the SM value for
$\kappa_\mathrm{\scriptscriptstyle EW}$ could be inconsistent with the
SM UT fit result. Finally, NP could produce contributions to EWP
operators such that $C_{9,10}\sim C_{7,8}$, or give rise to new
$\Delta I=3/2$ operators that cannot be eliminated. In this case, one
would observe $\vert R^{0,\mp} \vert \neq 1$. Present data give $\vert
R^{0} \vert = 0.96 \pm 0.17$. A small $\vert R^{0,\mp} \vert-1$ could
also be generated by $\langle Q_-\rangle$ or by a large dynamical
enhancement of $\langle Q_{7,8}\rangle$ within the SM.

\section{Conclusions and outlook}

We have presented a new method to constrain the Unitarity Triangle
using $B \to K \pi \pi$ decays.  This can be achieved with both
neutral and charged $B$ decays, using the amplitude ratios $R^0$ and
$R^\mp$ defined in eqs.~(\ref{eq:rn}) and (\ref{eq:rc}). The
theoretical uncertainty is negligible with respect to the foreseen
experimental error. We have discussed in detail how to take into
account electroweak penguins.  Our exploratory study shows that this
new constraint can have a sizable impact on the Unitarity Triangle
analysis in the near future. We have discussed possible improvements
and sensitivity to New Physics.

We thank the BaBar Collaboration for providing us with the results
published in ref.~\cite{babarkpp0} in the format needed for our
analysis.  We also thank G.~Cavoto, D.~Dujmic and R.~Faccini for
discussions and the careful reading of the manuscript. We are indebted
to J. Charles for finding a mistake in our original discussion of
electroweak penguins. We also thank M. Gronau for useful discussion on
the $SU(3)$ treatment of EWP's.  This work has been supported in part
by the EU network ``The quest for unification'' under the contract
MRTN-CT-2004-503369.

\end{document}